# Tight-Binding Superconducting Phases in the Unconventional Compounds Strontium-Substituted Lanthanum Cuprate and Strontium Ruthenate

Pedro Contreras[1, *], Dianela Osorio[1, 3], Eugeniy Yurievich Beliayev[2]

[1]Department of Physics, University of Los Andes, Mérida, Venezuela
[2]Verkin Institute for Low Temperature Physics and Engineering, National Academy of Sciences of Ukraine, Kharkiv, Ukraine
[3]Department of Brain and Behavioral Sciences, University of Pavia, Pavia, Italy

**Email address:**
pedrocontre@gmail.com (P. Contreras), daobfisica@gmail.com (D. Osorio), beliayev@gmail.com (E. Y. Beliayev)
[*]Corresponding author

**Abstract:** We use the idea of the Wigner probability distribution (WPD) in a reduced scattering phase space (RPS) for the elastic scattering cross-section, with the help of a Tight-Binding (TB) numerical procedure allowing us to consider the anisotropic quantum effects, to phenomenologically predict several phases in these two novel unconventional superconductors. Unlike our previous works with pieces of evidences that these two compounds are in the unitary strong scattering regime and that superconductivity is suppressed by the atoms of strontium in both materials, several phases are built. In the case of the strontium-substituted lanthanum cuprate, it was found three phases from one family of Wigner probabilistic distributions, one corresponding to the antiferromagnetic compound $La_2CuO_4$ another one which consists of a coalescing metallic phase for very lightly doped $La_{2-x}Sr_xCuO_4$, and finally a strong self-consistent dependent strange metal phase with optimal levels of doping. In the case of a triplet superconductor strontium ruthenate, three phases can be differentiated from two families of Wigner distribution probabilities, one family of WDP with point nodes where Cooper pairs and dressed scattered normal quasiparticles are mixed for the whole range of frequencies and which correspond to a FS γ-flat-sheet in the ground metallic state, and two phases from another WPD family, where, in one of then, the Miyake-Narikiyo quasinodal tiny gap model allows the unique presence of Cooper pairs in a tiny interval of frequencies near the superconducting transition $T_C$, the other phase corresponds to the mixed phase with Cooper pairs and dressed by stoichiometric strontium non-magnetic atoms, where strong self-consistent effects are noticeable. This approach allows comparing experimental results for samples in both compounds with numerical analysis studies.

**Keywords:** Strontium-Substituted Lanthanum Cuprate, Strontium Ruthenate, Unitary Limit, Tigh-Binding Non-magnetic Disorder, Elastic Scattering Cross-Section, Reduced Phase Space

## 1. Introduction

The reduced phase space (RPS) is a physical self-consistent distribution probability space obtained from the real and imaginary parts of the elastic scattering cross-section. It serves in physical phenomena where the unitary strong scattering regime takes place, such as in some unconventional superconductors, and where the mean quasiparticles free path can be approximate the following way: $l \sim a \sim k_F^{-1} \sim 1$.

In addition, the imaginary and real part of the elastic scattering cross-section, resembles the "Phase Space Distribution in Quantum-Mechanics" introduced by Wigner in 1932 [1, 2], it provides a classical window to look at some problems in the scattering theory of unconventional superconductors where the quantum phenomena are difficult to explain due mainly to the self-consistent frequency field effect. This idea was masterfully pointed out in a review written several years ago [2].

The idea of using the tight-binding methodology [3] with first neighbor hopping, including self-consistency due to non-magnetic disorder to study the RPS is expressed in Figure 1. The original idea of using the parametrization based on the inverse of the strength of the potential $c = (\pi N_F U_0)^{-1}$, and the concentration of impurities $\Gamma^+ = n_{imp}(\pi^2 N_F)^{-1}$ comes from [4]. However, the authors used an isotropic Fermi Surface and studied in particular one type of HTSC.

The first experiments investigating the influence of transition metal impurities in HTSCs, showed that nonmagnetic disorder suppress superconductivity [5, 6], on the contrary to BCS superconductors, where only magnetic impurities reduce the transition temperature $T_c$.

On the other hand, in $La_{2-x}Sr_xCuO_4$, the superconducting gap corresponds to a paired singlet scalar state $\Delta(k) = \Delta(-k)$ with line nodal behavior. The order parameter on the Fermi surface corresponding to the one-dimensional irreducible representation $B_{1g}$ of the tetragonal point symmetry group $D_{4h}$ [7]. The tight-binding anisotropic model corresponds to the nearest neighbor hopping expression for a band sheet centered at the corners $(\pm\pi/a, \pm\pi/a)$ of the first Brillouin zone in 2D case, $\xi(k_x, k_y) = \epsilon_F + 2t[\cos(k_x a) + \cos(k_y a)]$ with the TB anisotropic parameters $(t, \epsilon) = (0.2, 0.4)$ meV and an electron-hole (eh) symmetry $\xi_e(k_x, k_y) = \xi_h(-k_x, -k_y)$. The order parameter corresponds to the supercoducting gap $\Delta(k_x, k_y) = \Delta_0 \phi(k_x, k_y)$ with an irreducible representation basis function $\phi(k_x, k_y) = [\cos(k_x a) - \cos(k_y a)]$ and the zero gap parameter $\Delta_0 = 33.9$ meV [8, 9].

In addition, the other superconductor we address in this work concerns the family of ruthenates, particularly the unconventional triplet strontium ruthenate superconductor with multi-sheet (α, β and γ) FS [10]. For the triplet $Sr_2RuO_4$ [11] we use the expression corresponding to the Miyake – Narikiyo model [12] but with an anisotropic $\xi(k_x, k_y) = \epsilon_F + 2t[\cos(k_x a) + \cos(k_y a)]$.

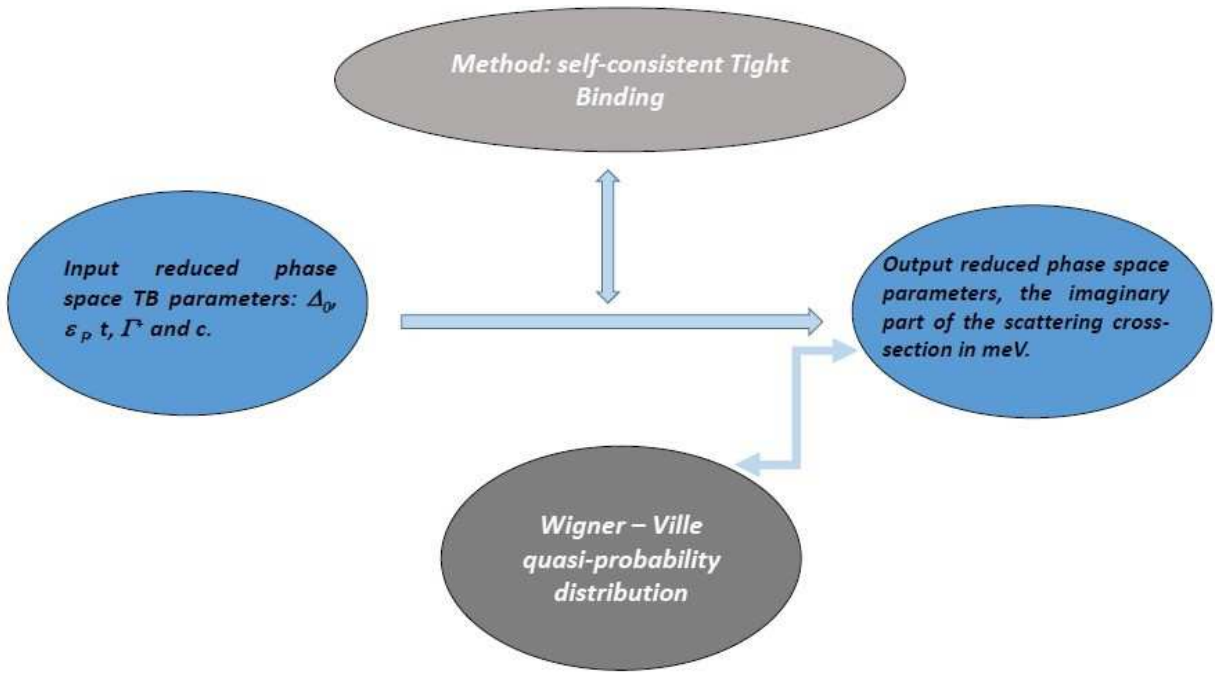

*Figure 1.* *The general scheme of the numerical procedure used. On the left side of the diagram are the input TB anisotropic parameters, the self-consistent numerical procedure, which includes several codes, is represented on the top. The output is given as the reduced scattering space in terms of WPD.*

$Sr_2RuO_4$ has TB anisotropic parameters of two kinds. First for the quasinodal model with a tiny gap $(t, \epsilon_F) = (0.4, -0.4)$ meV for a triplet state in the FS γ-sheet [13]. The second set, for a triplet state with point nodes $(t, \epsilon_F) = (0.4, -0.04)$ meV for a metallic ground state in the FS flat γ-sheet [14].

For both set of parameters $\Delta^\gamma(k_x, k_y) = \Delta_0 d^\gamma(k_x, k_y)$, with the vector $d^\gamma(k_x, k_y) = [(\sin(k_x a) + i\sin(k_y a)]\hat{z}$ and $\Delta_0^\gamma = 1.0$ meV, since there are experimental reports which estimated the value of $\Delta_0^\gamma$ to be less than 1 meV in impurity samples [15].

The nine points where the order parameter (OP) $d^\gamma(k_x, k_y)$ has zeros are, 4 points symmetrically distributed in the {10} and {01} planes at k-points $(0,\pm\pi)$ and $(\pm\pi,0)$, 4 points symmetrically distributed in the {11} planes at k-points $(\pm\pi,\pm\pi)$, and 1 point in the {00} plane at k-point $(0,0)$.

Following group theory considerations, the imaginary OP has two components that belong to the irreducible representation $E_{2u}$ of the tetragonal point group $D_{4h}$. It corresponds to a triplet odd paired state $d^\gamma(-k_x, -k_y) = -d^\gamma(k_x, k_y)$ with the basic functions $\sin(k_x a)$ and $\sin(k_y a)$ and Ginzburg-Landau coefficients $(1, i)$ [16].

Our contribution to this work is the study of three families of Wigner distribution probabilities aiming at numerically showing the different phases based on the disorder concentration $\Gamma^+$ phenomenological parameter. The impurity scattering disorder studies in the unitary limit for the elastic scattering cross-section (ECS) in a self-consistent manner is given by the real and imaginary parts of $\tilde{\omega}(\omega + i\, 0^+)$ and are plotted by the pair of points $(\Re(\tilde{\omega}), \Im(\tilde{\omega}))$ with tight-binding anisotropic 2D parameters. TB anisotropic modeling allowed us to successfully fit experimental low temperature data in two unconventional multiband superconductors at very low temperatures [17-23] for the triplet compound $Sr_2RuO_4$, and [24-28] for the HTSC $La_{2-x}Sr_xCuO_4$.

The computational and mathematical details of the algorithm were tested and reported for an isotropic Fermi surface and its corresponding order parameter in the previous work [29] for comparison with the seminal work by Carbotte and collaborators [4].

Figure 1 shows the general scheme of the numerical procedure used. On the left side of the diagram are the input TB anisotropic parameters, the self-consistent numerical procedure, which includes several codes, is represented on the top. The output is given as the reduced scattering space set of points $(\Re(\tilde{\omega}), \Im(\tilde{\omega}))$, which is related to the Wigner – Ville probabilistic distribution for non-magnetic disorder in the unitary limit. Those sets of 2D points are a product of the elastic scattering cross-section analysis since only energy is conserved in the unitary regime.

Since our primary goal is to visualize and describe briefly some properties of the three family of Wigner probabilistic distributions, we ought to emphasize that the whole analysis relays on the calculation of the function [9].

$$\tilde{\omega}(\omega + i\, 0^+) = \omega + i\, \pi\, \Gamma^+ \frac{1}{g(\tilde{\omega})},$$

Where the real part is given by $\Re(\tilde{\omega}) = \omega$, and the imaginary part is given by $\Im(\tilde{\omega}) = \pi\, \Gamma^+ \frac{1}{g(\tilde{\omega})}$.

The structure of this paper is as follows. In section 2, we present the first family of Wigner probabilistic distribution as a function of the impurity disorder concentration $\Gamma^+$, in the unitary limit for strontium-substituted lanthanum cuprate and present a table with the phases found by us in a recent work [28]. Section 3 presents the second and the third families of Wigner probabilistic distributions as a function of the impurity disorder concentration $\Gamma^+$ and the Fermi energy $\epsilon_F$ in $Sr_2RuO_4$.

The three families of Wigner distribution probabilities are in the unitary limit for the superconducting doped strontium-substituted lanthanum cuprate [28] and the strontium ruthenate with an extended analysis of the phases found in recent work [14]. The universal behavior was observed experimentally and studied theoretically at very low temperatures in for strontium ruthenate [12, 30]. The pioneering work for HTSC [4] was tested with a new self-consistent algorithm in [29].

For reviews on whether a gap with nodes can explain experimental data on anisotropic HTSCs and triplet superconductors, please see the anisotropic Fermi liquid formulation by MB Walker [31], and also a recent review by A. J. Leggett, and Y. Liu for $Sr_2RuO_4$ in [32]. The theory of non-magnetic scattering in normal metals is well elaborated by J. M Ziman [33]. The Edwards technique for unconventional superconductors is explained by V. Mineev and K. Samokhin [34].

For a comprehensive monograph on the self-consistent disorder problem in metals and alloys, see the book by I. M. Lifshitz, S. A. Gredeskul and L. A. Pastur [35]. For the quantum mechanical analysis of the elastic scattering cross-section check the monograph by L. Landau and E. Lifshitz [36]. The use of the idea of quasiparticles and its decay in solid state metallic systems is widely discussed by I. M. Lifshitz and M. I. Kaganov [37].

## 2. Wigner Probabilistic Distribution for Strontium-Substituted Lanthanum Cuprate

In this section, we present and visualize three phases in the compound $La_{2-x}Sr_xCuO_4$ in agreement with experiments of Dalakova and collaborators previously cited. The phases were built upon several plots for the pair points $(\Re(\tilde{\omega}), \Im(\tilde{\omega}))$ obtained from the elastic scattering cross-section. Since the calculation is in the unitary scattering limit, the peak at zero frequencies will coexist for all phases, contrasting its zero frequency value and the width of the Wigner probabilistic distribution as a function of the non-magnetic disorder concentration $\Gamma^+$. The prediction of the constant lifetime from the scattering cross-section was inferred in [38], the physical explanation has been given very recently in [39], and was compared successfully with experimental data in [28].

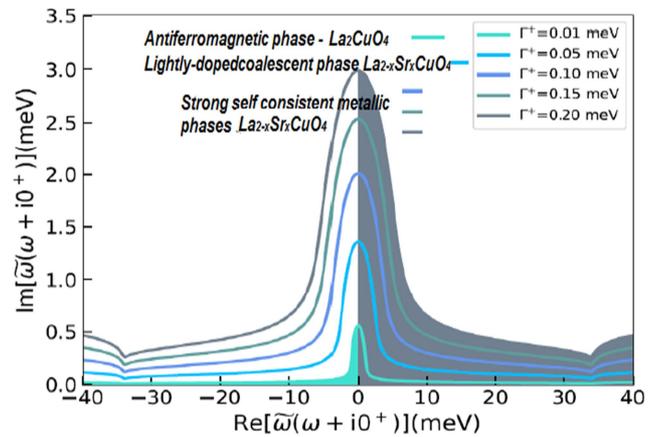

***Figure 2.*** *Plot of a family of WDP with Im $\tilde{\omega}(\omega +i\, 0^+)$ as a function of Re $\tilde{\omega}(\omega +i\, 0^+)$ for the $La_{2-x}Sr_xCuO_4$·HTSC.*

Therefore according to Figure 2, the clear blue curve is very close to the Mott - Hubbard dielectric $La_2CuO_4$ antiferromagnetic phase where there are no dressed quasiparticles since, with $\Im(\tilde{\omega}) \to 0$ and correspond to an insulator phase ($\Gamma^+ = 0.01$ meV). The blue plot corresponds to a metallic coalescent region with $\Im(\tilde{\omega}) \to const$ in a lightly doped $La_{2-x}Sr_xCuO_4$, (($\Gamma^+ = 0.05$ meV) and the other three plots correspond to the strange metal phase in doped $La_{2-x}Sr_xCuO_4$. Some data from the analysis of the plots are also presented in Table 1:

**Table 1.** *Phases and disorder parameters dependency for $La_{2-x}Sr_xCuO_4$.*

| Phases | Mott-Hubbard dielectric $La_2CuO_4$ | Light-doped coalescent $La_{2-x}Sr_xCuO_4$ | Strange metal phase |
|---|---|---|---|
| $\Gamma^+$ (meV) | 0.01 | 0.05 | 0.10-0.20 |
| Width $Im$ ECS (meV) | ~ 2 | ~ 4 | ~10 |

The physical explanation and experimental data for table 1 can be found in [28]. We remember that the model corresponds to Scalapino OP line nodes in the RPS.

## 3. Wigner Probabilistic Distribution for a Flat γ - Sheet with Point Nodes

In this section, we present and visualize one phase in $Sr_2RuO_4$. This phase was built by setting up to the ground state of the triplet superconductor, tuning the TB parameter $|\epsilon_F| = 0.04$ MeV, for the FS γ-sheet in $Sr_2RuO_4$ with several plots with pair of points $(\Re(\tilde{\omega}), \Im(\tilde{\omega}))$ obtained from the elastic scattering cross-section analysis as in the previous section for the singlet compound, and finally varying the non-magnetic disorder concentration $\Gamma^+$ parameter from the very dilute to more optimal values.

The calculation in the unitary scattering limit for stoichiometric Sr, the width at zero frequencies will be stronger than for the doped LaCuO ceramic [13], changing the zero frequency value and the width of the Wigner probabilistic distribution as a function of $\Gamma^+$. See table 2 for the values.

The point nodes behavior corresponds to a FS γ-flat sheet in the metallic ground state, and the gap touches the FS in 4 points symmetrically distributed in the {10} and {01} planes at k-points $(0, \pm\pi)$ and $(\pm\pi, 0)$, see [14] for a complete theory for this case. It resembles the Scalapino nodal lines OP with smaller scattering window in meV. It can distinguish in Figure 3 one phase, with point nodes shadowed blue for small impurity concentration, and another phase shadowed gray for optimal values of non-magnetic dirt, that strongly mixes both types of quasiparticles, bosons and dressed normal quasiparticles.

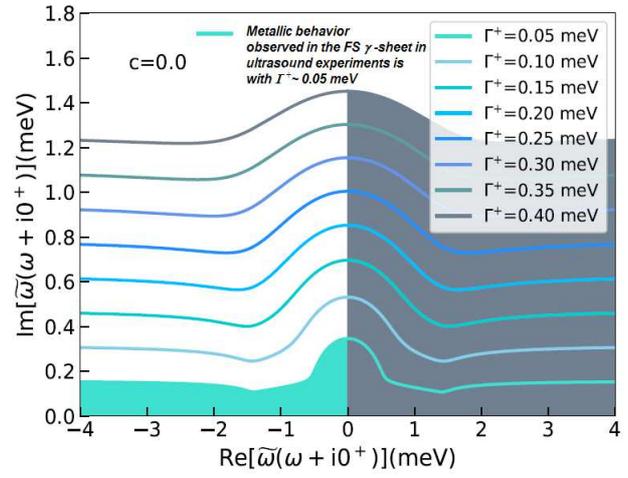

**Figure 3.** *Plot of a family of WDP with $Im\ \tilde{\omega}(\omega + i0^+)$ as a function of $Re\ \tilde{\omega}(\omega + i0^+)$ for the ground state flat γ band of $Sr_2RuO_4$.*

**Table 2.** *Ground state phase and disorder parameters dependency for a flat point nodes γ sheet $Sr_2RuO_4$.*

| One Phase | Metallic behavior with almost constant quasiparticles lifetime | Metallic behavior with stronger stoichiometric Sr and still const. dressed lifetime |
|---|---|---|
| $\Gamma^+$ (meV) | 0.05 | 0.10 – 0.40 |
| Width $Im$ ECS (meV) | ~ 0.5 | ~ 0.6 – 0.9 |

## 4. Wigner Probabilistic Distribution for Quasi-Point Triplet Nodes

This section presents and visualizes another two phases in $Sr_2RuO_4$. These phases were built by going to the state with quasi-point nodes of the Miyake Narikiyo model, making the tight-binding parameter $|\epsilon_F| = 0.4$ meV corresponding to the same order of magnitude that the hoping parameter "t" for the FS γ-sheet shown in Figure 4.

Each WPD family has several plots with pairs $(\Re(\tilde{\omega}), \Im(\tilde{\omega}))$ taken from the elastic scattering cross-section as in the previous sections, varying the non-magnetic disorder concentration $\Gamma^+$. In Figure 4, we can distinguish two phases, one the Miyake Narikiyo phase with only Cooper pairs shadowed blue (($\Gamma^+ = 0.05$ meV).), that correspond to a tiny MN interval with only boson quasiparticles and another phase shadowed gray (($\Gamma^+ >= 0.10$ meV).) that mixes both types of quasiparticles, bosons and dressed normal quasiparticles [13, 14, 39].

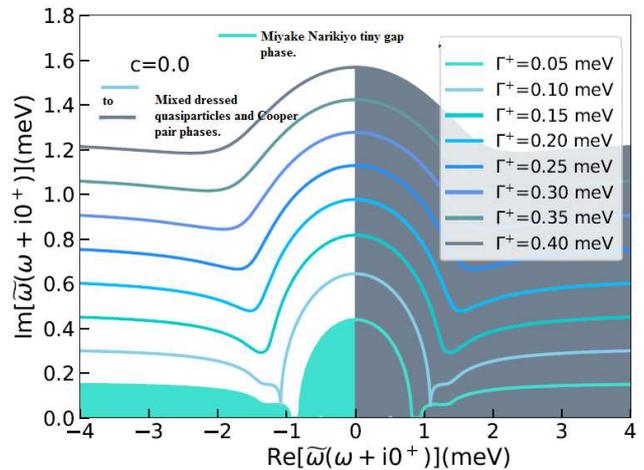

**Figure 4** *Plot of a family of WDP with $Im\ \tilde{\omega}(\omega + i0^+)$ as a function of $Re\ \tilde{\omega}(\omega + i0^+)$ for the MN $Sr_2RuO_4$ model.*

The calculation is in the the unitary scattering limit for stoichiometric Sr with point nodes and a tiny gap, the width at zero frequencies will be stronger than for the strontium doped LaCuO ceramic and the previous ground state family of plots for stoichiometry $Sr_2RuO_4$, increasing its zero frequency values and the width of the Wigner probabilistic distributions as a function of the non-magnetic disorder concentration $\Gamma^+$, as can bee seen in Table 3.

We think that, further computational studies with advanced computer high performance tools using the the tigh-binding approach of Prof. Harrison [3] could give a better understanding of these two phases by understanding the width of the resonance, henceforth expanding our simplistic approach of Figure 1.

*Table 3. Summarizes the two phases predicted in [29].*

| Two Phase | Metallic behavior with only Cooper pairs, the Miyake-Narikiyo tiny gap bosons phase | Metallic behavior with stronger stoichiometric Sr and mixed Cooper pairs and dressed quasiparticles. |
|---|---|---|
| $\Gamma^+$ (meV) | 0.01 | 0.05 – 0.40 |
| Width *Im* SCS (meV) | ~ 1.8 | ~ 2.2 – 2.8 |

## 5. Conclusions

The present work aimed to investigate and summarize the behavior of the elastic scattering cross-section for two unconventional superconductors, the singlet strontium-substituted lanthanum cuprate and the triplet strontium ruthenate as a function of the concentration of non-magnetic disorder in a self-consistent tight-binding approach as summarized in Figure 1.

The use of self-consistent energy analysis for $\tilde{\omega}(\omega + i\, 0^+)$ in Planck units proves to be very useful ($\hbar = k_B = c = 1$) for realistic 2D anisotropic unconventional superconductors with line nodes, quasi-point nodes and points nodes to compare with experiments, since it outputs several families of Wigner distribution probabilities that help to differentiate numerically a set of phases in these two compounds.

Some interesting physical properties such as coalescence for very low amount of non-magnetic dirt, and pure Bose supercarriers and mixed dressed scattered quasiparticles with Cooper pairs phases can be phenomenological differentiate in this study in the strong scattering unitary regime [28, 39].

This approach resembles the findings of the liquid isotope $^3$He in the 1970s [40], where several phases were found experimentally and theoretically in the superconductor state of its liquid state [41] as it is sketched in figure 5.

Particularly we would like to emphasize that the supercoducting triplet phase of strontium ruthenate with quasi-point nodes has several wonderful similarities with the supercoducting A phase of the liquid isotope $^3$He, where in both cases, an external field is able to split the time reversal symmetry of a single phase due to symmetry breaking external field effects, a wonderful and unique physical effect manifested at macroscopic level [41-44]. See [45] for a pedagogical explanation of Figure 5.

In addition, the TB Fermi surface parameters and value gap consistent with experimental ARPES data gives a better insight into the experimental problem by opening a classical window to the quantum phenomena in these two superconductors, despite one is a HTCS ceramic and the other is a low temperature superconductor, the approach representation through the WPD analysis becomes a very useful tool in this both cases.

Further studies following this approach include the flat FS case for HTSC and other line nodes OP in unconventional superconductors.

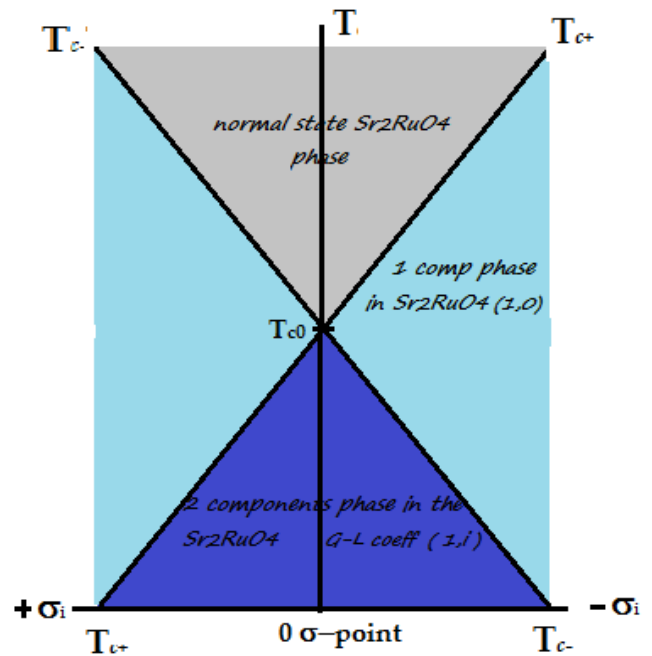

*Figure 5. Symmetry time reversal breaking field effects in $Sr_2RuO_4$ due to uniaxial stress is similar to the breaking of the specific heat in the liquid isotope $^3$He A-phase, in the superconducting phase A (picture from our work [45]). Both materials (the solid and liquid A component) are in a triplet OP superconducting phase.*

## 6. Recommendations

The use of the reduced scattering space (RPS) for dressed by nonmagnetic impurities quasiparticles as a function of the disorder and strength parameters $\Gamma^+$ and c, seems to be a powerful tool when it is combined with the analysis of the elastic cross-section in unconventional superconductors.

An RPS self-consistent analysis of the elastic cross-section allows the identification of different phases, such as the metallic coalescent phase with an almost constant lifetime in HTSC, the strange metallic phase with a strong self-consistent dependence on previous energy values, or the tiny gap inhomogeneous phase.

In addition, the irreducible representation of some points crystal groups combined with a few TB parameters provide useful insights in these types of studies, therefore it is advisable to test other irreducible representations and other TB parameters for other unconventional superconductors, such as different HTSC or heavy fermions. It could help to explain some of the physical properties of these alloys. Finally, the study of flat band materials as we did for $Sr_2RuO_4$ will occupy definitely in oncoming years a new topic for computational material design, please see the book by J. Shaginyan and M. Amusia [46].

## Acknowledgements

This research did not receive any specific grant from funding agencies in the public, commercial, or not-for-profit sectors. Also we thank an anonymous reviewer whose technical comments helped improve and clarify this manuscript. Finally, we acknowledge the effort of PG Science Publishing Group in the mission of promoting academic exchanges and the advancement of science by helping the most disadvantaged scientists around the world.